\documentclass[paper,twocolumn,twoside]{geophysics}

\usepackage{hyperref}
\usepackage{amsmath}
\usepackage{amsfonts}

\usepackage{balance}

\DeclareMathOperator*{\argmax}{arg\,max}

\begin{document}

\renewcommand{\figdir}{Images} 
\newcommand{\xs}[1]{\mathbf{x}_{#1}}

\title{Extracting small time-lapse traveltime changes in a reservoir using primaries and internal multiples after Marchenko-based target zone isolation}

\renewcommand{\thefootnote}{1} 

\ms{} 

\address{
Delft University of Technology, Department of Geoscience and Engineering, Delft, The Netherlands. 
}
\email{
\href{mailto:J.E.vanIJsseldijk@tudelft.nl}{J.E.vanIJsseldijk@tudelft.nl} (corresponding author); J.R.vanderNeut@tudelft.nl; J.W.Thorbecke@tudelft.nl; C.P.A.Wapenaar@tudelft.nl. \\
}
\author{Johno van IJsseldijk\footnotemark[1], Joost van der Neut\footnotemark[1], Jan Thorbecke\footnotemark[1] and Kees Wapenaar\footnotemark[1]}

\footer{}
\lefthead{van IJsseldijk et al.}
\righthead{Time-lapse changes with the Marchenko method}
\maketitle

\sloppy

\begin{abstract}
\hspace{.4cm}Geophysical monitoring of subsurface reservoirs relies on detecting small changes in the seismic response between a baseline and monitor study. However, internal multiples, related to the over- and underburden, can obstruct the view of the target response, hence complicating the time-lapse analysis. In order to retrieve a response that is free from over- and underburden effects, the data-driven Marchenko method is used. This method effectively isolates the target response, which can then be used to extract more precise time-lapse changes. Additionally, the method also reveals target-related multiples that probe the reservoir more than once, which further define the changes in the reservoir. To verify the effectiveness of the method, a numerical example is constructed. This test shows that when using the isolated target response, the observed time differences resemble the expected time differences in the reservoir. Moreover, the results  obtained with target-related multiples also benefit from the Marchenko-based isolation of the reservoir. It is, therefore, concluded that this method has the potential to observe dynamic changes in the subsurface with increased accuracy.
\end{abstract}

\section{Introduction}
\hspace{\parindent}Time-lapse seismic studies are concerned with detecting small changes in the seismic response between a baseline and a monitor study. These changes can either be a difference in amplitude \citep[e.g.][]{Landro2001}, a difference in traveltime \citep[e.g.][]{Landro2004} or a combination of both \citep[e.g.][]{Trani2011}. These time-lapse methods are essential for observing and monitoring subsurface reservoirs, with applications ranging from determining pressure and fluid saturation changes \citep{Landro2001} to monitoring CO$_2$ injection \citep{Roach2015} or observing compaction in a reservoir \citep{Hatchell2005}. 

In order for these methods to work optimally, it is important that the reservoir response can be clearly identified in the seismic response. In practice, this requirement is not always fulfilled, as multiple reflections from a (highly) reflective overburden can mask the response of the reservoir. It is, therefore, desirable to remove the overburden effects before applying any time-lapse analysis. The Marchenko method is able to redatum a wavefield from the surface of the earth to an arbitrary focal depth in the subsurface while accounting for all orders of multiples \citep{wapenaar2014marchenko,slob2014seismic}. This data-driven method can be used to remove all interactions from layers above the selected focal level, hence giving an unobstructed view of the reservoir response. From this new response, the traveltime difference in the reservoir can more precisely be determined.

In addition to removing the overburden, the reservoir response can completely be isolated by also removing the underburden with the Marchenko method \citep{WapenaarAndStaring2018}. Consequently, not only the primary response of the reservoir is uncovered, but also internal multiples, which traversed through the reservoir more than once, will now also be clearly visible and unobstructed by primaries and multiples outside the target zone. Since these multiples have passed through the reservoir multiple times, the time-lapse traveltime \ifthenelse{\boolean{@manu}}{}{\pagebreak} change of the multiples will be larger, hence more sensitive. This is akin to coda-wave interferometry, which exploits the fact that time-lapse changes are exaggerated in the coda due to the longer paths traveled in the medium \citep{Snieder2002,Gret2005}. 

Inspired by this principle of coda-wave interferometry, \citet{WapenaarIJsseldijk2020} show how correlation of multiples improve the ability to detect small changes in velocity compared to correlation of primaries. This method is then adapted to find changes in lateral varying media \citep{vanIJsseldijk2021}. In this work we further develop the method in order to account for time-lapse changes in the overburden. First, we revise the theory of isolating the reservoir response with the Marchenko method and review how to extract traveltime changes from this isolated response. Furthermore, we show how multiples traveling through the reservoir can be enhanced, in order to improve the accuracy of the retrieved time shifts. Subsequently, we present a numerical model that will be used to test the methodology. The reservoir response is then isolated from the modeled data, and the traveltime changes of the primary as well as the multiple reflections are calculated. Finally, we discuss the results and possible future improvements to the method.

\section{Theory}

\hspace{\parindent}Time-lapse seismic experiments aim to resolve the differences between a baseline study at time $t_1$ and a monitor study conducted at a later time $t_2$. These differences can be attributed to changes inside of a reservoir and overburden, for example due to production and geomechanical processes. Here we propose a method by which the reservoir response is isolated separately for the baseline and monitor studies, after which cross-correlation between the two studies is used to find the traveltimes differences. 

\autoref{fig:Principle} shows the principle of the proposed method. Here the acoustic situation is considered, with a reservoir enclosed by two strong reflectors. Note how the primary from the first reflector, does not probe the reservoir, whereas the primary from the second reflector does. Moreover, the internal multiples generated by these reflectors will traverse the reservoir multiple times, hence they experience a larger traveltime shift. In order to achieve this same situation from a regular reflection response measured at the surface, the medium is first divided into 3 parts: overburden "a", target zone "b" which contains the reservoir and two reflectors as in \autoref{fig:Principle}, and underburden "c". The reflection response of the full medium is denoted by $R_{abc}(\xs{R},\xs{S},t)$, here $\xs{R}$, $\xs{S}$ and $t$ denote the receiver position, source position and time, respectively. Our first aim is to isolate the reflection response $R_{b}$ of the target zone  with the help of the Marchenko method, which will briefly be discussed in the next section.

\plot[tb!]{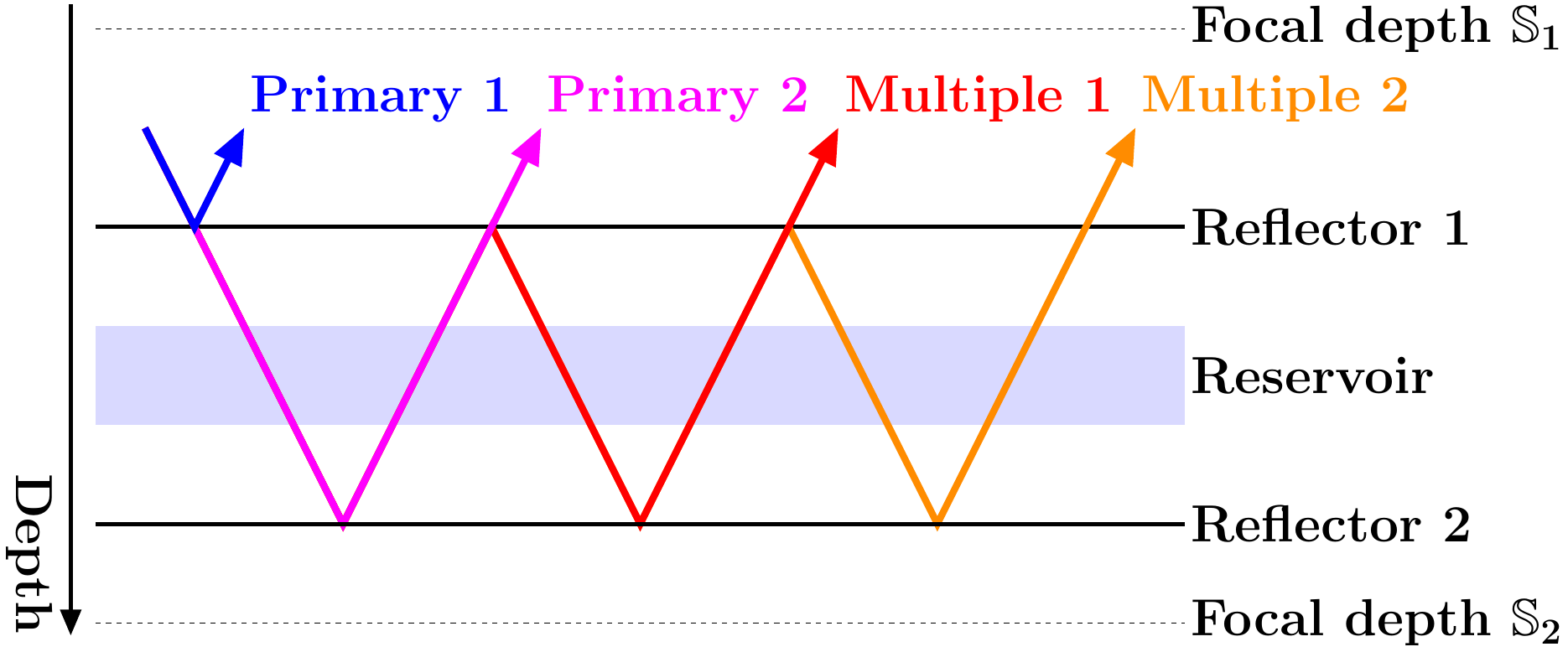}{width=.47\textwidth}
{Graphic displaying the principle of the method. Note how the reservoir layer is located in between two reflectors. Primary 1 from reflector 1 does not propagate through the reservoir, whereas primary 2 from reflector 2 does. The multiples (1 and 2) are propagating through the reservoir twice or thrice, hence experiencing double or triple the traveltime changes compared to primary 2. Target zone "b" is located in between focal depths $\mathbb{S}_1$ and $\mathbb{S}_2$, overburden "a" and underburden "c" are above and below the target zone, respectively.}

\subsection{Extrapolated Marchenko representations}

\plot*{MarchenkoFunctions}{width=\textwidth}
{Schematic representation of the extrapolated focusing- and Green's functions. (A) The focusing functions $f^+_1$ (in blue) and $f^-_1$ (in red) are defined in a medium truncated below the focal level ($\mathbb{S}_F$). They are extrapolated to the surface ($\mathbb{S}_0$) using the direct arrival of the transmission response $T_d$ (in orange) to create the extrapolated focusing functions $v^\pm$. (B) the downgoing Green's function ($G^{-,+}$) is extrapolated to create $U^{-,+}$. (C) the upgoing Green's function ($G^{-,-}$) creates it's extrapolated counterpart $U^{-,-}$.} 

\hspace{\parindent}At the base of the Marchenko method are the focusing functions ($f_1^\pm$) that allow for retrieval of the Green's functions ($G^{-,\pm}$) between the acquisition surface $\mathbb{S}_0$ and a focal level in the subsurface. Here the left superscript $-$ denotes that the wavefield is upgoing at the receiver position (at the focal level) and the right superscript $\pm$ denotes a down- or up-going direction from the source position (at the acquisition surface). \Citet{vanderNeut2016} introduce modified functions that are extrapolated to the surface by convolution with the direct arrival of the transmission response ($T_d$). These extrapolated focusing functions ($v^\pm$) are defined as follows:
\begin{equation}
\label{eqn:focii}
v^\pm(\xs{R},\xs{S}',t) = \int_{\mathbb{S}_F} f_1^\pm(\xs{R},\xs{F},t) * T_{d}(\xs{F},\xs{S}',t) d\xs{F}.
\end{equation}
Here $\xs{F}$ is the coordinate of the focusing point at focal depth $\mathbb{S}_F$, $\xs{S}'$ is a coordinate on the acquisition surface, and $*$ denotes temporal convolution. Note, that except for $\xs{F}$, all coordinates in this work refer to positions at the surface $\mathbb{S}_0$. Similarly, the extrapolated Green's functions ($U^{-,\pm}$) are defined as:
\begin{equation}
\label{eqn:greens}
U^{-,\pm}(\xs{R},\xs{S}',t) =  \hspace*{4cm}
\end{equation}
\begin{equation*}
\hspace*{1cm} \int_{\mathbb{S}_F} G^{-,\pm}(\xs{R},\xs{F},t) * T_{d}(\xs{F},\xs{S}',\pm t) d\xs{F}.
\end{equation*}
These two equations are visualized in \autoref{fig:MarchenkoFunctions}, which shows how the extrapolated functions are related to the original focusing and Green's functions. By using these extrapolated functions the retrieved wavefields derived in the next section will be situated at the surface $\mathbb{S}_0$ and not at focal level $\mathbb{S}_F$ as is the case with the regular Marchenko functions.
Finally, the same convolutions are applied to the coupled Marchenko representations to find the extrapolated representations:
\begin{equation}
\label{eqn:mar1}
U^{-,+}(\xs{R},\xs{S}',t) + v^-(\xs{R},\xs{S}',t) = \hspace*{2cm}
\end{equation}
\begin{equation*}
\hspace*{1cm} \int_{\mathbb{S}_0} R(\xs{R},\xs{S},t) * v^+(\xs{S},\xs{S}',t)d\xs{S},
\end{equation*}
\begin{equation}
 \label{eqn:mar2}
U^{-,-}(\xs{R},\xs{S}',-t) + v^+(\xs{R},\xs{S}',t) =  \hspace*{2cm}
\end{equation}
\begin{equation*}
\hspace*{1cm} \int_{\mathbb{S}_0} R(\xs{R},\xs{S},-t) * v^-(\xs{S},\xs{S}',t)d\xs{S}.
\end{equation*}
The reflection response is denoted by $R$. In this paper, this response will either be the response of the full medium $R_{abc}$ or the response after overburden removal $R_{bc}$. Moreover, these two equations have four unknowns. In order to solve this system a causality constraint is introduced, which exploits the fact that the focusing and Green's functions are separable in time. In order to apply this constraint, an estimate of the two-way travel time between the focal level and the surface is required. In our case, this is achieved by computing the direct arrival of the Green's function in a smooth velocity model with an Eikonal solver, and then convolving this response with itself to find the two-way travel time. A more elaborate derivation of the Marchenko method is beyond the scope of this paper. Instead, the reader is referred to \citet{Wapenaar2021}, who give more background on both the regular and extrapolated expressions.

\plot[tb!]{flowchart1iso}{width=.47\textwidth}
{Flowchart depicting how the reservoir response is isolated with the Marchenko method.}

\subsection{Isolation of the reservoir's response}

\hspace{\parindent}Using the relations presented in the previous section, the focusing and Green's functions above and below the reservoir can now be retrieved. From these functions, the reflection response of the target zone can be isolated. First, the overburden is removed, using the extrapolated Green's function between the overburden and the upper boundary $\mathbb{S}_1$ of the target zone \citep{Wapenaar2021}:
\begin{equation}
\label{eqn:redatum}
U^{-,+}_{a|bc}(\xs{R},\xs{S}',t) = \hspace*{4cm}
\end{equation}
\begin{equation*}
\hspace*{1cm} -\int_{\mathbb{S}_0} U^{-,-}_{a|bc}(\xs{R},\xs{R}',t) * R_{bc}(\xs{R}',\xs{S}',t) d \xs{R}'.
\end{equation*}
Here $U^{-,\pm}_{a|bc}$ are the extrapolated Green's functions, retrieved with the Marchenko method from equations \ref{eqn:mar1} and \ref{eqn:mar2}, where $R_{abc}$ is used as reflection response $R$. The vertical line in the subscript indicates the location of focal level, i.e. between the overburden "a" and the target-zone "b". \autoref{eqn:redatum} is solved for $R_{bc}$ by multidimensional deconvolution \citep[MDD,][]{Wapenaar2011b}. This MDD is achieved with least-squares inversion in the frequency domain. Effectively we have now acquired a new reflection response $R_{bc}$, which is free from overburden interactions. Furthermore, coordinates $\xs{R}'$ as well as $\xs{S}'$ are located at the surface, due to the use of the extrapolated Green's functions. In contrast, previous work with regular Green's functions acquired a redatumed response at the focal depth, and then required an additional step to extrapolate this response to the surface \citep{vanIJsseldijk2021}.

Next, the newly acquired reflection response ($R_{bc}$) is used to retreive the extrapolated focusing functions between the target zone and the upper boundary $\mathbb{S}_2$ of the underburden. These focusing functions are then used to remove the underburden: 
\begin{equation}
\label{eqn:isolation}
v^-_{b|c}(\xs{R},\xs{S}',t) = \hspace*{5cm}
\end{equation}
\begin{equation*}
 \int_{\mathbb{S}_0} v^+_{b|c} (\xs{R},\xs{R}',t) * R_{b}(\xs{R}',\xs{S}',t) d \xs{R}'.
\end{equation*}
The subscript $b|c$ denotes that $R_{bc}$ was used to retrieve the focusing functions from equations \ref{eqn:mar1} and \ref{eqn:mar2}, with the focal level between target zone "b" and underburden "c". Note, that \autoref{eqn:isolation} directly follows from the definition of the focusing functions in the truncated medium \citep{WapenaarAndStaring2018}. Again, the reflection response of the target zone $R_b$ can be resolved from \autoref{eqn:isolation} by means of MDD. 

\plot[tb!]{flowchart2ccs}{width=.47\textwidth}
{Flowchart to get the time differences in the reservoir from the isolated response (continuation of \autoref{fig:flowchart1iso}).}

Effectively, the target zone response has now been isolated, leaving a response analogous to the situation in \autoref{fig:Principle}, but with the sources and receivers at the surface $\mathbb{S}_0$.
As a final step the multiples in the final response $R_{b}$ can be further amplified. First, consider that the multiples in $R_{b}$ continue infinitely in time, and are constructed from the focusing functions $v^-_{b|c}$ and $v^+_{b|c}$, which are finite in time (i.e. they are confined between $t=0$ and the two-way travel time to the focal depth). Next, $v^+_{b|c}$ in \autoref{eqn:isolation} is divided into an initial function $v^+_{b|c,0}$ and a coda $v^+_{b|c,m}$ \citep{Wapenaar2021}:
\begin{equation}
\label{eqn:vdecomp}
v^+_{b|c} (\xs{R},\xs{R}',t) = \hspace*{5cm} 
\end{equation}
\begin{equation*}
\delta (\xs{H,R}-\xs{H,R}') \delta (t) +  v^+_{b|c,m} (\xs{R},\xs{R}',t).
\end{equation*}
Here, $\delta$ denotes the Dirac delta function. From this equation it follows that the initial function $v^+_{b|c,0}$ can be interpreted as a (bandlimited) delta pulse at $t=0$. This pulse is followed by the coda $v^+_{b|c,m}$. 
Appendix A shows, when solving \autoref{eqn:isolation} for $R_b$, that $v^+_{b|c,0}$  is mainly responsible for the primaries in $R_{b}$, whereas $v^+_{b|c,m}$ updates these primaries and constructs the subsequent multiples of the response. Therefore, by amplifying $v^+_{b|c,m}$ the multiples in response $R_{b}$ should get enhanced as well. Note, this enhancement will cause the amplitudes of the response not to be accurate anymore. However, this is not an issue for the current implementation, since only time-shifts are desired. 
\autoref{fig:flowchart1iso} shows an overview of the process to isolate the target response. In this chart only a smooth velocity model of the baseline is used. It is assumed that this model can also be used for the monitor study, because the velocity changes are relatively small and only an approximation of two-way traveltime to the focal depth is needed. Next, the new responses $R_b$ for the baseline and monitor will be used to extract the traveltime shifts in the reservoir.

\subsection{Extraction of time differences}

\plot*{Models}{width=\textwidth}
{Velocity (A) and density (B) model of the baseline study for the numerical example. The black dashed lines define the focal levels above ($\mathbb{S}_1$) and below the reservoir ($\mathbb{S}_2$), used for the Marchenko method. The solid white contour depicts the three different reservoir pockets. C shows the difference in velocity between the baseline and monitor study. The density inside the reservoir pockets is also increasing with $100 \ kg/m^3$ for the monitor study, with no density changes outside the reservoir. Primary 1 and Primary 2 originate from the green to blue contrast at $700m$ to $900m$ and the blue to green contrast at $1000m$, respectively.}

\hspace{\parindent}Before extracting the traveltime shifts in the reservoir, the different primaries and multiples are identified. Primary 1 and 2 are easily detected due to the isolation of the target zone (i.e. there are no interactions from the overburden to obscure the primaries). Subsequently, the arrival times of the internal multiples can be approximated based on the primaries, where the arrival time of the $n$-th multiple can be approximated by the arrival time of primary 2 plus $n$ times the difference in time between the two primaries. The first step is now to eliminate any time shifts resulting from a time-lapse change in the overburden. In order to do this the temporal correlation between primary 1 (i.e. the response that is not penetrating the reservoir) and primary 2 or an internal multiple (i.e. the responses that go through the reservoir) is computed. This gives the correlation between primary 1 and the target response below the reservoir:
\begin{equation}
\label{eqn:correlation}
C_{*}(\mathbf{x}_0,\tau) = \hspace*{6cm} 
\end{equation}
\begin{equation*}
\int_0^\infty \Theta_{P1}(t+\tau) R_{b}(\mathbf{x}_0,t+\tau) \Theta_{*}(t) R_{b}(\mathbf{x}_0,t) d t.
\end{equation*}
Here, $C$ is the correlation of the two responses, and $\xs{0}$ denotes the position of the zero-offset traces in the data, where $\xs{S} = \xs{R}$. This correlation contains the time-lag between the first reflector and P1, M1 or M2. Theta is a time window or mute function that isolates a specific primary or multiple as follows: 
\begin{equation}
\label{eqn:theta}
\Theta_{*}(t) = \begin{cases}
1, \text{if } t_{*}-\epsilon < t \leq t_{*} + \epsilon \\
0, \text{otherwise.}
\end{cases}
\end{equation}
In these equations the asterisk $*$ can be replaced with P1, P2, M1 or M2, for primary 1, primary 2, multiple 1 and multiple 2, respectively. Hence $t_*$ is the travel time of one of the primaries or multiples. $\epsilon$ is a small constant that defines the window, and makes sure the whole waveform is included. Any traveltime differences in the overburden are removed by first calculating the time-lag with primary 1 in \autoref{eqn:correlation}, i.e. the time difference between primary 1 and either P2, M1 or M2 is free from overburden interactions. After the time-lags of \autoref{eqn:correlation} have been independently calculated for the baseline and monitor study, the time-lapse traveltime shifts in the reservoir can be determined with a second temporal correlation as follows:
\begin{equation}
\label{eqn:extraction}
\Delta t_{*}(\mathbf{x}_0) = \hspace*{6cm}
\end{equation}
\begin{equation*}
 \argmax_\tau \left( \int_0^\infty C_{*}(\mathbf{x}_0,t+\tau) \bar{C}_{*}(\mathbf{x}_0,t) d t \right).
\end{equation*}
The bar denotes that the monitor correlation is used, thus the time-lags, of primary 2 or one of the multiples with respect to primary 1, for the baseline and monitor are correlated. Next, the argument of the maximum of this correlation is taken to determine the traveltime differences in the reservoir. The process of extracting the time shifts is summarized in \autoref{fig:flowchart2ccs}. Note that there are a few additional practical considerations included in the chart, such as resampling and removing outliers. These will be discussed in more detail in the next section.

\section{Numerical example}

\hspace{\parindent}A numerical experiment is designed, to show the viability of the method. The baseline velocity and density models are shown in figures \ref{fig:Models} A and B, respectively. \autoref{fig:Models} C displays the change in velocity for the monitor model. In the overburden, there is a velocity decrease of $25\ m/s$, whereas the velocity and density in the three reservoir compartments increase by $100\ m/s$ and $100\ kg/m^3$. The reflection responses of the baseline ($R_{abc}$) and monitor ($\bar{R}_{abc}$) are computed with finite differences,  using a wavelet with a flat spectrum between 5 and 80 Hz \citep{thorbecke2011finite}. The receivers are placed along a 6000$m$ long line with a spacing of 10$m$, and the data are recorded with a sampling rate of 4$ms$. The 601 sources are excited at the same positions as the receivers. Estimates of the two-way traveltimes, between the surface and the focal points at 675$m$  and 1100$m$, are acquired using an eikonal solver in a smooth version of the baseline velocity model of \autoref{fig:Models}A. Lastly, a band-limited delta pulse is computed, which is used as the initial focusing function ($v_0^+$) in the iterative Marchenko scheme. 

\subsection{Results of target zone isolation}

\plot*[tb!]{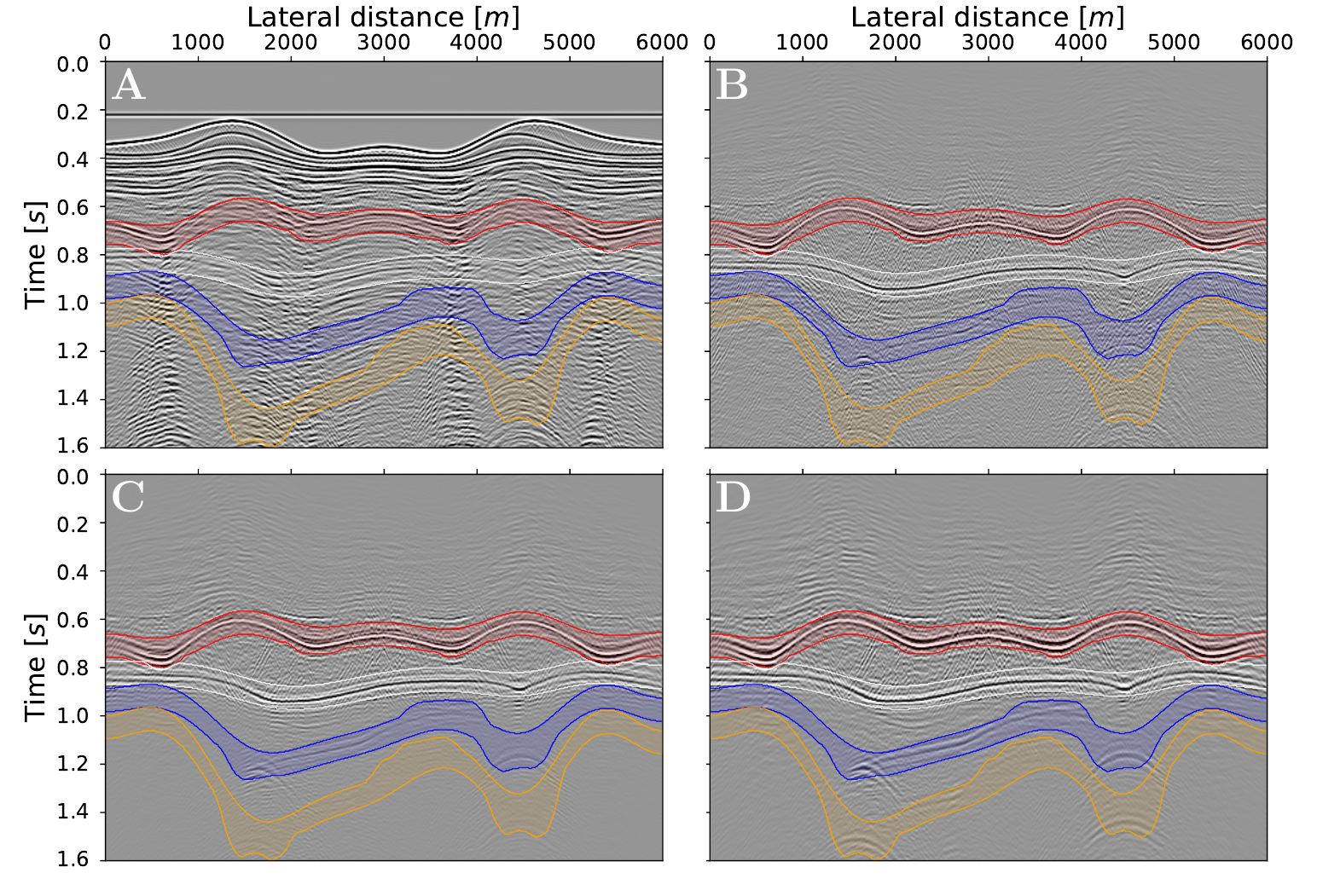}{width=\textwidth}
{Zero-offset gathers for the baseline reflection response of the entire medium $R_{abc}$ (A), after overburden removal $R_{bc}$ (B), after over- and underburden removal $R_{b}$ (C), after isolation and multiple enhancement of $R_{b}$ (D). The windows used for cross-correlation for primary 1, primary 2, multiple 1 and multiple 2 are highlighted in red, white, blue and orange, respectively. }

\hspace{\parindent}\autoref{fig:Windows}A shows a zero-offset section of the initial reflection response before applying the multiple internal removal ($R_{abc}$). Note that only the zero-offset data are shown, but the data at all offsets are available and used to compute the isolated reservoir response $R_b$ from $R_{abc}$. Due to the highly reflective overburden, the primaries (P1 in red and P2 in white) are nearly impossible to identify, and multiple 1 and 2 (in  blue and orange) are completely obscured by the overburden and underburden interactions. After removing the overburden and acquiring $R_{bc}$ (\autoref{fig:Windows}B), the primaries are now clearly recognizable in the seismogram. However, the windows enclosing the multiples contain undesirable events from primary reflections from the underburden. The third panel (\ref{fig:Windows}C) reveals that these events are successfully removed after underburden removal. Also, note how the multiple events are enhanced in the fourth display (\ref{fig:Windows}D). This is the result of the previously described scaling factor applied to $v^+_{b|c,m}$ (which was scaled with a factor 2.5) before retrieving $R_b$. 

All the time windows that select the primaries and multiples in \autoref{fig:Windows} are picked from this final response in panel D ($R_{b}$). First, the arrival time for both primaries is manually selected. As stated before, these times are then used to provide an estimate for the multiple arrivals. Finally, the windows are manually adjusted to ensure they include the full response from each individual event. In the next section, these windows will be used for the cross-correlations that compute the time differences.

\subsection{Time differences inside the reservoir}

\plot*[tb!]{results_scl1}{width=.95\textwidth}
{Results showing the estimated time difference in the reservoir from primary 2 (A), multiple 1 (B) and multiple 2 (C), no scaling factor has been applied to enhance the multiples. First, the time-lag with primary 1 is computed for the baseline and monitor study. These time-lags are then cross-correlated to find the time differences. In each plot the blue line shows the result derived from the full reflection response ($R_{abc}$), the orange and green lines are the time difference derived from $R_{bc}$ and $R_b$, respectively. The red line shows the cross-correlations of the ideal data, where $R_b$ is computed using finite differences. The light blue show the change in time for 1D zero-offset traces (i.e. two times the reservoir thickness times the difference in the slowness).}

\hspace{\parindent}From the isolated response, the traveltime changes can be estimated. First, the data are interpolated from $4ms$ to $1ms$ to achieve a better time resolution. Second, the primary enclosed in the red window is selected from the data. Similarly, either the second primary, first multiple or second multiple is also extracted using its respective window (as shown in \autoref{fig:Windows}). For both the baseline and monitor studies, these responses are then correlated to effectively remove the time differences developed in the overburden, i.e. this correlation retrieves the time-lag between P1 and P2, M1 or M2, thus removing the shared path in the overburden. Finally, the baseline and monitor time-lag correlations are mutually correlated to find the time differences in the target zone. 

The results of these final correlations are shown in \autoref{fig:results_scl1}. These result can be interpreted as the zero-offset time-lag differences between the baseline and monitor surveys. Hence any deviation from $\Delta t = 0$ should represent the time shift within the reservoir. Since the velocity in the reservoir is increasing, a negative $\Delta t$ is expected, whereas a positive shift indicates a velocity decrease (i.e. with the current model positive $\Delta t$ would indicate that the result is contaminated with remaining overburden effects). Note that these results were acquired with the true Marchenko scaling without the previously described multiple enhancement. Here, the response of the full medium, the response after overburden removal and the response of the target zone (i.e. after over- and underburden removal) are displayed with blue, orange and green lines, respectively. The light-blue area marks the 1D-based zero-offset traveltime difference, which was computed by multiplying two times the reservoir thickness with the slowness change in the reservoir. However, this reference solution does not take into account lateral variations. The red line gives a second reference solution made by cross-correlating "ideal" data. This data was acquired in a model with a smooth overburden (A), the same target zone (B) as the actual model and a transparent underburden (C). The zero-offset response is then modeled with finite-differences, providing an accurate isolated response of the target zone $R_b$ for both the monitor and baseline response. Subsequently, the primaries and multiples are identified in the modeled zero-offset response, and then correlated as described in flowchart in \autoref{fig:flowchart2ccs}. 

\autoref{fig:results_scl1}A shows the results for primary 2. Although, all three responses capture some differences in the reservoir, the response of the full medium still reflects changes in the overburden as shown by the time-shifts larger than $0ms$. These positive time-shifts are almost fully removed after overburden removal. Note that, on the one hand, the correlations do not match the 1D reference very well, because of the lateral variations in the model. On the other hand, the match with correlation of the ideal data (the red line) is a lot better, which implies that the Marchenko based isolation was successful.
Based on these results it is concluded that the expected time differences are smaller than $15ms$ for P1, M1 and M2. This observation is used to achieve more accurate results, by removing outliers that give a time difference larger than $15ms$ at any lateral distance (i.e. they are removed before applying the Gaussian smoothing along the lateral direction in \autoref{fig:flowchart2ccs}).

Next, the procedure is applied to multiple 1 and multiple 2, the results of which are shown in figures \ref{fig:results_scl1}B and \ref{fig:results_scl1}C, respectively. This time none of the results match either reference perfectly, and seemingly no meaningful information can be acquired from the multiple 2. However, these results are achieved without any multiple enhancement. In the next section, we will explore how the results can be improved by using multiple enhancement. 

\subsection{Results after multiple enhancement}

\hspace{\parindent}To improve results, i.e. to get more accurate time-differences, the multiple enhancement is now applied, by scaling $v^+_{b|c,m}$ with a factor of 2.5. This factor was chosen somewhat arbitrarily, but as a rule of thumb the maximum amplitude in the new $v^+_{b|c,m}$ should not exceed 80\% of the maximum amplitude in $v^+_{b|c,0}$. The results of the correlations of this new $R_b$ are shown in \autoref{fig:results}. The time differences acquired by correlation with primary 2 (\autoref{fig:results}A) show no significant differences from the original results, but the correlations of the multiples of the isolated response (the green lines in \ref{fig:results}B and \ref{fig:results}C) match the "ideal" data (the red lines) a lot better now. 
On the contrary, when looking at the results for the multiple 1 (\autoref{fig:results}B), a clear dissimilarity is observed between the results of the isolated response in green and the full response in blue. Furthermore, note the improvement relative to results obtained with primary 2 at $4500m$ lateral distance, where the correlation matches the 1D reference a lot better. The same can be seen in the results of multiple 2 in \autoref{fig:results_scl1}C. The two other reservoir compartments at $1500m$ and $3000m$ do not present the same improvements, instead, their results confirm the observations for the correlations with primary 2. 

Especially, the results for primary 2 and multiple 1 (the green lines in \ref{fig:results}A and \ref{fig:results}B) accurately match the correlation of the "ideal" data. The results of multiple 2 do not have the same match. This is due to the fact that events from multiple 1 are interfering within the correlation window of multiple 2 for the "ideal" data, and leaving an imprint on the reference solution. From this match with the "ideal" data, it can be concluded that the Marchenko method succeeded in correctly isolating the target response, as the results for multiple 1 coincide with the reference solution. This also highlights the importance of isolating the response, since the correlations of the multiples in $R_{abc}$ and $R_{bc}$ do not come close to the reference solution at all. Finally, the results outside the reservoir compartments should show a time differences equal to $0ms$, and time differences larger than $0ms$ indicate that the result is contaminated by effects from the overburden.
The correlation results of the isolated response $R_b$ display less of these positive time differences compared to the results of the full response $R_{abc}$. Consequently, this is another confirmation that the isolation process has successfully eliminated the effects of the overburden.

\plot*[tb!]{results}{width=.95\textwidth}
{Same results as \autoref{fig:results_scl1}, but with multiple enhancement by applying a scaling factor of 2.5 to $v^+_{b|c,m}$ before retrieving $R_b$ with MDD. Note that the multiple correlations of the isolated response $R_b$ (the green lines in \ref{fig:results}B and \ref{fig:results}C) are a lot closer to the ideal result (the red lines) compared to the results without multiple enhancement (figures \ref{fig:results_scl1}B and \ref{fig:results_scl1}C).}

\section{Discussion}

\hspace{\parindent}Although the results show that the reservoir response can successfully be isolated and accurate time differences can be retrieved from both the primary reflection and the internal multiples, there are a number of issues that require a more in-depth discussion. In addition to this discussion, future improvements and practical considerations for the implementation on real data will be considered as well.

First, the scaling factor to $v^+_{b|c,m}$, introduced to amplify multiple events, was found experimentally. 
The authors conducted numerous 1D experiments to get a better understanding of the effect of the scaling factor, and found that the scaled $v^+_{b|c,m}$ should not exceed the maximum amplitude in $v^+_{b|c,0}$, in order to maintain a stable result, with as a rule of thumb the amplitudes preferably staying below 80\% of this maximum. Due to these nuances it is always advised to obtain the results without any multiple enhancement first, and only make an effort to improve the results with multiple enhancement after this initial result is achieved.

A second point of discussion is the order of operations used to isolate the target response. In theory, it does not matter whether the underburden is removed before or after overburden removal. Numerical experiments indeed showed that removing the underburden before the overburden is also a viable approach to isolate the target response, by first retrieving $R_{ab}$ from a MDD of $v^{\pm}_{ab|c}$ and then $R_b$ from $U^{-,\pm}_{a|b}$. However, the previously discussed multiple enhancements are no longer available when the method is applied in this order, because the MDD of the focusing functions would be applied before the MDD of the Green's functions that require proper scaling. Therefore, it was decided to start with overburden elimination followed by removal of the underburden. 

Third, the method is designed to use as little a priori information as possible, needing solely three prerequisites: the baseline reflection response ($R_{abc}$), the monitor reflection response ($\bar{R}_{abc}$) and a smooth version of the baseline velocity model. The smooth velocity model is used to approximate the two-way travel time between the surface and the focal depth. The same model can be used for both the baseline and monitor reflection response because it is assumed that the velocity changes in the medium are relatively small.  When the velocity changes are large, a separate velocity model is required to isolate the target zone from the monitor reflection response, but the application of the method would not change otherwise.

In this work the time shifts are retrieved by a simple cross-correlation method. Instead of cross-correlation, a waveform-based or other method could be used to possibly improve the accuracy of the time shift results further. \citet{RP4Dshifts} give a comprehensive overview of the different available methods to calculate time shifts.

Next, the primary results before the target zone isolation already match the actual results quite closely, and only minor deviations due to overburden effects are present (i.e. the parts where $\Delta t \ > \ 0ms$). However, these results indirectly benefit from the isolated result, because the windows, which are used to identify the primaries, are selected from the isolated results (\autoref{fig:Windows}C). Moreover, previous results by \citet{vanIJsseldijk2021} show that 
the correlations from $R_{abc}$ are insufficient to get accurate time differences in less complex models with overburden events interfering with the primaries.

Application of the Marchenko algorithm to field data can be quite cumbersome. Especially the MDD that is used to remove the overburden effects tends to be very sensitive to errors in the amplitude of the data. To overcome this limitation, \citet{staringetal2018} introduce a double-focusing method, which is more stable but leaves some remaining interactions of the overburden. A similar approach is envisioned to acquire $R_{bc}$, when applying this method to real data. It is noted, however, that any errors in $R_{bc}$ will affect the final result of $R_b$ as well. Moreover, the Marchenko method is quite sensitive to wrong amplitudes in field data. In order to overcome this limitation either a scaling factor can be determined using cost functions \citep{Brackenhoff2016} or  an advanced 3D to 2D conversion can be applied \citep{DukalskiReinicke2022}.

Finally, we would ideally find the velocity change of the reservoir rather than the time differences. For very simple situations a similar approach as coda-wave interferometry can be considered, which finds the velocity perturbation from the change in traveltime and initial velocity \citep{snieder2006theory}. However, this only holds when the relative velocity perturbation is constant at every location. In our case, the perturbation is different outside the reservoir, hence the relation does not hold. Alternatively, the velocity perturbation can be found by inversion of cross-correlations at all offsets (instead of just the zero-offset data used here). Compared to the traveltime differences, the velocity changes can more directly be related to physical processes such as flow in the reservoir. Currently, this is subject to ongoing research. 

\section{Conclusion}

\hspace{\parindent}A good understanding of fluid flow, temperature variations and mechanical changes in subsurface reservoirs is essential for a large variety of geoscientific methods. These dynamic changes can be observed with seismic time-lapse methods by identifying amplitude changes, time shifts or both, between a baseline and a monitor study. However, the response of a subsurface target can be obscured by interferences from reflectors in the overburden and/or underburden, making it harder to detect the time-lapse effects. The Marchenko method can be used to remove primaries as well as internal multiples above or below an arbitrary focal level in the subsurface from the reflection response. Hence, this method can be used to isolate the reservoir response in both the baseline and monitor response, enabling an unobstructed examination of changes in the target zone. 

A twofold methodology has been proposed to extract time differences. With this methodology, first the target response is isolated, by overburden removal, followed by underburden removal. This new response is then used to identify the primary and multiple reflections in the target zone. Second, time differences are retrieved by cross-correlating the different reflections of the baseline and monitor studies. By first correlating the response with primary 1 above the reservoir, all possible time shifts of the overburden are removed. 

A numerical example with a strongly reflective overburden was created to test the methodology. The isolation of the target zone revealed the primary responses of the reservoir, allowing their extraction from the data. Next, the time differences of the reservoir could be approximated from correlations with a primary reflection below the reservoir. Furthermore, additional traveltime changes were retrieved from the first and second-order multiples, created by the two reflectors enclosing the reservoir. These multiples confirmed the earlier observations, but also improved specific blind spots in the original approximation of the time changes.

The proposed methodology provides a new means to extract traveltime differences, especially for situations where complex overburden and underburden interactions mask the target response. Future developments should also make it possible to invert for velocity changes in the reservoir, rather than time differences. The method will then enable us to more accurately observe dynamic changes in the subsurface.

\begin{acknowledgments}
\hspace{\parindent}The authors are grateful for the insightful comments of three anonymous reviewers. We also thank Leon Diekmann, Giovanni Meles, Myrna Staring and Joeri Brackenhoff for insightful discussions about this work. This research was funded by the European Research Council (ERC) under the European Union’s Horizon 2020 research and innovation programme (grant agreement no. 742703).
\end{acknowledgments}

\begin{data}
\hspace{\parindent}Data associated with this research are available and can be accessed via the following URL: \href{https://gitlab.com/geophysicsdelft/OpenSource}{https://gitlab.com/geophysicsdelft/OpenSource}. The figures in this paper can be reproduced using the demo in the ``vmar" folder of the repository.
\end{data}

\append{Multiple Enhancement}
\label{app:multiples}

This appendix aims to give a more comprehensive explanation as to why increasing the weight of $v^+_{b|c,m}$, results in enhanced multiples in the final reflection response $R_b$. First operator $\mathcal{V}_m^+$ is defined to apply a multidimensional convolution with $v^+_{b|c,m}$, as follows: 
\begin{equation}
\mathcal{V}_m^+ R_b(\xs{R},\xs{S}',t) = \hspace*{4cm} 
\end{equation}
\begin{equation*}
\int_{\mathbb{S}_0} v^+_{b|c,m} (\xs{R},\xs{R}',t) * R_{b}(\xs{R}',\xs{S}',t) d \xs{R}'.
\end{equation*}
Next, this equation and \autoref{eqn:vdecomp} are substituted in \autoref{eqn:isolation}. After rearranging the terms, an expression to obtain $R_b$ is acquired:
\begin{equation}
\label{eqn:rbinv}
R_{b}(\xs{R},\xs{S}',t)= \left(1+\mathcal{V}_m^+\right)^{-1} v^-_{b|c}(\xs{R},\xs{S}',t).
\end{equation}
Finally, the inverse in \autoref{eqn:rbinv} can be expanded as a Neumann series:
\begin{equation}
\label{eqn:rbneu}
R_{b}(\xs{R},\xs{S}',t) = \sum_{k=0}^\infty \left(-\mathcal{V}_m^+ \right)^k v^-_{b|c}(\xs{R},\xs{S}',t).
\end{equation}
\autoref{eqn:rbneu} illustrates how the primaries and multiples in $R_b$ are constructed from the focusing functions. First, for $k=0$ only the contributions from $v^-_{b|c}$ are available. This will account for the primaries and multiples contained in $R_b$ from times $t=0$ up to times below the two-way traveltime to the focal depth, since $v^-_{b|c}$ is equal to zero outside this range. Consequently, the multiples at times larger than the two-way traveltime have to be constructed from terms with $k>0$, which will be created using $v^+_{b|c,m}$. 
This implies that the multiples in $R_b$ can artificially be enhanced by amplifying $v^+_{b|c,m}$. Application of \autoref{eqn:rbneu} is only stable if the L2 norm of the operator is less then 1, i.e. $||\mathcal{V}_m^+||^2_2 < 1$. This constraint does not necessarily hold for all subsurface models. Because of this limitation the Neumann series is solely introduced here to provide an intuitive explanation for the multiple enhancement, whereas the MDD in \autoref{eqn:isolation} is solved by least-squares inversion in the frequency domain.


\bibliographystyle{seg}  
\bibliography{TL_bib}

\balance

\end{document}